\RequirePackage{amsmath}
\documentclass[runningheads]{llncs}

\usepackage{comment}
\usepackage[utf8]{inputenc}
\usepackage[T1]{fontenc}
\usepackage[english]{babel}
\usepackage{microtype}
\usepackage{paralist}
\usepackage{silence}
\usepackage{caption}
\usepackage{subcaption}

\usepackage[
    backend=biber,
    bibstyle=numeric,
    citestyle=numeric-comp,
    sorting=nty,
    sortcites=true,
    maxbibnames=20,
    urldate=comp,
]{biblatex}

\usepackage[svgnames,x11names]{xcolor}
\usepackage{hyphenat}
\usepackage{tabularx}

\usepackage{graphicx}
\usepackage{xspace}
\usepackage{amsmath,amssymb,amsfonts,amsthm}
\usepackage{pifont}
\usepackage{multirow}
\usepackage{makecell}
\usepackage{multicol}

\usepackage{float}
\usepackage{adjustbox}
\usepackage{booktabs}
\usepackage{wasysym}
\usepackage{xstring}
\usepackage{soul}
\usepackage[shortlabels,inline]{enumitem}
\setlist[enumerate]{(1)}
\usepackage[
    group-separator={,},
    group-minimum-digits=4,
    detect-weight=true,
]{siunitx}
\usepackage[english=american,strict]{csquotes}
\MakeOuterQuote{"}
\usepackage{balance}
\usepackage[nomessages]{fp}
\usepackage{caption}
\usepackage{listings}
\usepackage[colorlinks=true,urlcolor=black]{hyperref}
\usepackage{cleveref}
\Crefname{lstlisting}{Listing}{Listings}

\PassOptionsToPackage{hyphens}{url}
\hypersetup{
    colorlinks,
    linkcolor={black},
    citecolor={red!70!black},
    urlcolor={blue!70!black}
}
\Crefname{lstlisting}{Listing}{Listings}

\usepackage[disable]{todonotes}
\makeatletter%
\if@todonotes@disabled%
\else%
\fi
\makeatother

\usepackage{listings}

\definecolor{lstgreen}{rgb}{0,0.6,0}

\lstdefinestyle{plain}{%
  numbers=none,
  frame=none,
  xleftmargin=1pt,
  xrightmargin=1pt,
}

\lstdefinelanguage{EDL}[ISO]{C++}{%
  morekeywords=[2]{user_check,in,out,size,string}%
}%

\lstdefinelanguage{Rust}{%
  morekeywords={as,break,const,continue,crate,else,enum,extern,false,
    fn,for,if,impl,in,let,loop,match,mod,move,mut,pub,ref,return,Self,
    self,static,struct,super,trait,true,type,unsafe,use,where,while,
    abstract,alignof,become,box,do,final,macro,offsetof,override,priv,
    proc,pure,sizeof,typeof,unsized,virtual,yield},
  morekeywords=[2]{isize,usize,char,bool,str,String,u8,u16,u32,u64,u128,i8,i16,i32,i64,i128,f32,f64},
  sensitive=true,
  morecomment=[l]{//},
  morecomment=[l]{///}, 
  morestring=[b]{"},
}%

\definecolor{rust_structs}{RGB}{10, 130, 84}
\definecolor{rust_calls}{RGB}{120, 0, 45}
\definecolor{rust_comment}{RGB}{1, 117, 15}

\lstdefinestyle{rust-style}{
    language=Rust,
    basicstyle=\scriptsize\ttfamily,
    columns=fullflexible,
    keepspaces=false,
    showstringspaces=false,
    showtabs=false,
    commentstyle=\color{rust_comment},
    keywordstyle=\color{blue}\bfseries,
    morekeywords={guard, assert},
    stringstyle=\color{black},
    classoffset=1,
    morekeywords={ProgramResult, BaseCommitmentHashingAccount, ElusivError, ComputationIsNotYetStarted, solana_program,sysvar,instructions, ProgramError,InvalidAccountData, Pubkey, AccountInfo,Ok,BTreeMap,MigrationFunction,Some},
    keywordstyle=\color{rust_structs},
    classoffset=2,
    morekeywords={compute_base_commitment_hash, get_is_active, compute_base_commitment_hash_partial, load_instruction_at, try_borrow_mut_data, map_err, next_wallet_account_info,iter,next_account_info,next_program_account_info, handle,collect_remaining_balance,migrations,from,get,borrow,borrow_mut,read_serialized_signer},
    keywordstyle=\color{rust_calls},
    classoffset=0,
    frame=tb,
    framesep=4pt,
    rulesepcolor=\color{gray},
    numbers=left,
    numberstyle=\tiny,
    xleftmargin=5mm,
    breaklines=true,
    breakatwhitespace=true,
}

\usepackage{tikz}
\usetikzlibrary{shapes,arrows,positioning,shapes.multipart}

\newcommand{\circleone}{\ding{192}\xspace}

\newcommand{\circletwo}{\ding{193}\xspace}

\newcommand{\sbpf}{sBPF\xspace}
\newcommand{\bn}{Binary Ninja\xspace}
\newcommand{\ssc}{Solana programs\xspace}
\newcommand{\ACPI}{Arbitrary Cross-Program Invocation\xspace}
\newcommand{\acpi}{arbitrary cross-program invocation\xspace}

\newcommand{\p}[1]{\SI{#1}{\percent}}

\newcommand{\gb}[1]{\SI{#1}{\giga\byte}}

\let\underscore\_
\renewcommand{\_}{$\underscore$\allowbreak{}}
\newcommand{\var}[1]{\textit{#1}\xspace}

\newcommand{\sym}[1]{\texttt{#1}\xspace}
\newcommand{\func}[1]{\texttt{#1}\xspace}

\newcommand{\yes}{$\checkmark$\xspace}
\newcommand{\no}{$\times$\xspace}

\newlength{\nowidthtmp}

\newcommand{\fds}{FuzzDelSol\xspace}

\newcommand\shortcon[1]{\texttt{\StrLeft{#1}{5}.\!.\!.\StrRight{#1}{3}}}

\renewcommand{\paragraph}[1]{\par\medskip\noindent\textbf{#1.}\hspace{1ex minus .1ex}}

\let\oldsubsubsection\subsubsection
\renewcommand\subsubsection[1]{\par\medskip\oldsubsubsection{#1}}

\newcommand\totalanalyzed{\num{8714}\xspace}
\newcommand\totalbugs{\num{467}\xspace} 

\newcommand{\settikzmark}[1]{%
  \tikz[overlay,remember picture,baseline] \coordinate (#1) at (0,0) {};}

\newcommand{\greenplus}{{\color{green} +}}

\newcommand{\tikzlinehighlight}[3]{%
  \draw[#1,line width=6pt,opacity=0.2]%
  ([yshift=1.5pt]#2) -- ([yshift=1.5pt]#3);%
}

\newcommand{\diffhighlightadd}[2]{%
  \tikzlinehighlight{YellowGreen}{#1}{#2}
}

\newcommand{\tool}{\textsc{SseRex}\xspace}
\newcommand{\tools}{\textsc{SseRex}'s\xspace}
\title{\tool: Practical Symbolic Execution of Solana~Smart~Contracts}
\author{
Tobias Cloosters\inst{1}\orcidID{\small 0009-0008-3682-0197}
\and Pascal Winkler\inst{1}\orcidID{\small 0009-0002-8324-4993}
\and Jens-Rene Giesen\inst{1}\orcidID{\small 0009-0004-0685-6237}
\and Ghassan Karame\inst{2}\protect\footnotemark[1]\orcidID{\small 0000-0002-2828-4071}
\and Lucas Davi\inst{1}\orcidID{\small 0000-0002-7322-2777}
}
\institute{
University of Duisburg-Essen, Essen, Germany. \email{\{firstname.lastname\}@uni-due.de}\\
\and Ruhr University Bochum, Bochum, Germany. \email{ghassan@karame.org}
\footnotetext[1]{The author was additionally affiliated with NEC Laboratories Europe at the time of writing.}
}
\date{}

\begin{document}
\usetikzlibrary{shapes,arrows,positioning,shapes.multipart}
\usetikzlibrary{backgrounds}
\usetikzlibrary{fit}
\usetikzlibrary{calc}

\maketitle

\begin{abstract}
Solana is rapidly gaining traction among smart contract developers and users.
However, its growing adoption has been accompanied by a series of major security incidents, which have spurred research into automated analysis techniques for Solana smart contracts.
Unfortunately, existing approaches do not address the unique and complex account model of Solana.
In this paper, we propose \tool, the first symbolic execution vulnerability detection approach for finding Solana-specific bugs such as missing owner checks, missing signer checks, and missing key checks, as well as \acpi{}s.
Our evaluation of \totalanalyzed~bytecode-only contracts shows that our approach outperforms existing approaches and identifies potential bugs in \totalbugs~different contracts.
Additionally, we analyzed 120 open-source Solana projects and conducted in-depth case studies on four of them. Our findings reveal that subtle, easily overlooked issues often serve as the root cause of severe exploits, further highlighting the need for specialized analysis tools like \tool.

\end{abstract}


\section{Introduction}

The Solana blockchain is gaining popularity due to its ability to provide smart contract functionality with high transaction throughput and low transaction fees~\cite{Pierro_2022_can,Yakovenko_2018_solana}.
Solana smart contracts, also called \emph{programs}, are usually written in Rust and compiled to an eBPF-based bytecode~\cite{solana_programs}.
Unlike the Linux kernel's eBPF, it lacks the control-flow verification mechanism~\cite{linux_ebpf}.
This lack of validation has led to numerous Solana exploits in recent years~\cite{solana_exploits_2022,cashio_hack,mangomarket_hack}, resulting in a loss of more than \$523~million.
The exploited contracts exhibited bugs ranging from logical errors, like missing input data validation~\cite{cashio_hack,mangomarket_hack}, to missing signature verification, as seen in the Wormhole exploit~\cite{wormhole_hack}.

Solana programs separate data from logic to employ concurrent transaction execution~\cite{Cardenas_2024_lollipop,solana_accounts}, requiring the full context state in every Solana transaction.
As this context is created by a potentially malicious creator of the transaction, there is a significant risk that such a context is manipulated.
To mitigate this, the Solana runtime provides trustworthy metadata---validated a~priori---such as public keys, account owners, account signatures, and cryptographically derived accounts---so-called program-derived addresses (PDAs).
While automated validation exists, effective validation remains highly dependent on the contract's specific goal.
As a remedy, the Anchor~\cite{anchor} framework employs Rust types and macros to automatically check accounts and metadata.
Yet, Anchor cannot cover the full space of attacks; developers must still manually perform signer checks and verify data dependencies such as authority ownership of a given wallet.

Recently, several approaches have been proposed to detect vulnerabilities in Solana.
VRust~\cite{Cui2022_nm} analyzes source code for specific bug patterns; however, source code is largely unavailable for Solana contracts.
Further, VRust suffers from a high false positive rate, e.g., only one of the 108~reported bugs was indeed a missing key check.
\fds~\cite{smolka_2023_fuzz} applies fuzzing to a custom Solana runtime to detect bugs during program execution.
Fuzzing fundamentally prioritizes speed and is limited by the high computational overhead required to implement effective oracles, and its sensitivity to ``magic bytes'' in the input---such as public keys and type identifiers---which are essential for exploring meaningful execution paths.
In general, current analysis approaches have a limited view of the execution context and cannot holistically evaluate a contract's code.
For instance, VRust's patterns need explicit adaptation to find errors in Anchor programs, whereas \fds cannot analyze Anchor contracts, which currently make up at least \p{43} of deployed contracts.

In this paper, we introduce \tool, the first \emph{symbolic execution framework} for vulnerability detection in \ssc.
Symbolic execution, by design, comprehensively captures data flows and program constraints~\cite{He_2019_learning,Su_2015_combining}.
As our approach demonstrates, it enables precise tracking of all account verifications and their relations, without sacrificing generality, as evidenced by its ability to analyze Anchor contracts (in contrast to previous work~\cite{Cui2022_nm,smolka_2023_fuzz}).
\tool counteracts path explosion via static pre-analysis to resolve account deserialization and string formatting.
Further, we develop hybrid symbolic lengths, which avoid symbolic pointers but allow symbolic data lengths.
We design specialized oracles for Solana-specific issues---missing key, owner, or signer checks---that evaluate whether an execution path leads to unexpected or unsafe behavior.
Particularly, they flag critical actions---such as cross-program invocations or state-modifying account writes---occurring without proper verification of the involved entities.

We comprehensively evaluate \tool on \emph{all} deployed Solana contracts, a dataset of \totalanalyzed~bytecode-only contracts, 120~projects with source code from the Anchor registry, and previous approaches' datasets~\cite{Cui2022_nm,smolka_2023_fuzz}.
Our results show that \tool significantly reduces false positives and discovers more vulnerabilities than previous work.
In particular, the large-scale study identifies \totalbugs~vulnerable contracts.
We performed intensive manual analysis of a sample of 30 reported bugs, and verified 26 as true bugs (\p{87} true positive rate).
Finally, we also present four detailed case studies to showcase the challenges of finding bugs in Solana.

To foster research on the security of \ssc, we will release all components of \tool publicly on GitHub.


\section{Background}

Solana uses an account-based model with stateless execution, where accounts store program bytecode or data~\cite{solana_account}.
Solana programs receive all data via instructions and cannot access any external state.
Programs commonly dispatch logic based on instruction bytes, as in the Anchor framework~\cite{anchor}.
The runtime provides metadata for each account---including owner, signer field, executable flag, and balance in lamports~(1\,bn lamports = 1\,SOL)~\cite{solana_2024_lamport}.
The \emph{signer field} indicates that the account signed the transaction~\cite{solana_transactions}.
Programs can interact via cross-program invocation~(CPI)~\cite{solana_2024_cpi} and delegate access using Program Derived Addresses~(PDAs) deterministically derived from program ID and arbitrary seed bytes~\cite{solana_2024_pda}.

\begin{figure}[t]
    \centering
    \includegraphics[width=0.9\linewidth, alt{
    An attacker generates a transaction that contains malicious data accounts, unsigned user accounts or malicious programs that result in different exploits in a program.
    A program is parented by the system program, and has PDA accounts and data accounts as children.
    }]{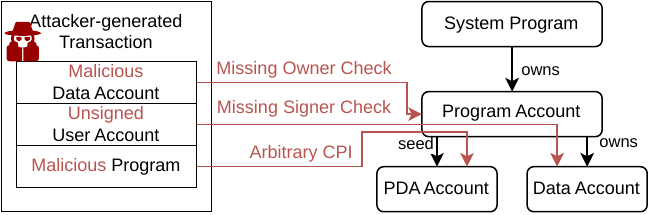}
    \caption{Malicious attack transactions in Solana and their targets.}
    \label{fig:solana_account_model}
\end{figure}

\subsection{Solana's Threat Model}

\begin{figure}[tb]
\centering
\begin{lstlisting}[escapechar=@, style=rust-style, caption={The contract exhibits multiple vulnerabilities that an attacker can exploit.}, label={lst:exploit_contract}]
pub fn process_instruction(program_id: &Pubkey, accounts: &[AccountInfo], ix_data: &[u8],) -> ProgramResult {
    let vault_account = &accounts.iter_mut().nth(0)?; // vault state [owner check]
    let authority_account = &accounts.iter_mut().nth(1)?; // admin [signer check]
    let target_program = &accounts.iter_mut().nth(2)?; // target program [acpi]
    let cpi_accounts: Vec<AccountInfo> = accounts_iter.cloned().collect(); // cpi args

    // MOC assert_eq!(vault_account.owner, program_id); @ \label{lst:moc_lines_start} @
    let data = vault_account.try_borrow_data()?;       // trusts unverified account
    let balance = u64::from(&data[0..8].into()); // data from unverified account @ \label{lst:moc_lines_end} @

    // MSC if !authority_account.is_signer { return Err(...) } @ \label{lst:msc_lines_start} @
    let expected_authority = ...; // often from another account
    if authority_account.key != &expected_authority {return Err(...);}
    // attacker supplies correct key without needing the private key @ \label{lst:msc_lines_end} @

    let cpi_ix = solana_program::instruction::Instruction { @ \label{lst:cpi_lines_start} @
        program_id: *target_program.key,   // unvalidated CPI target
        accounts: cpi_accounts, // attacker-controlled accounts
        data: ix_data.to_vec()};            // attacker-controlled CPI data
    invoke(&cpi_ix, &cpi_accounts)?;// attacker can manipulate program and accounts@ \label{lst:cpi_lines_end} @
// ...
\end{lstlisting}
\end{figure}

Since the creator of a transaction can freely select input data, programs must carefully validate inputs before performing critical actions, e.g., transferring funds, writing data to an account, or performing CPI~\cite{neodyme_2024_solanapitfalls}.
As shown in \Cref{fig:solana_account_model}, there are three strategies to manipulate the execution context.
First, an attacker may impersonate another user and authorize a transaction due to a missing signer check~(MSC)~\cite{slowmist_2024_solanasecurity,neodyme_2024_solanapitfalls,auditfirst_2022_solanavulnerabilities}.
For example, in \Cref{fig:solana_account_model}, the attacker supplies an unsigned user account and authorizes access to the data account of the impersonated user, as the program lacks an explicit signer check.
As illustrated in \Cref{lst:exploit_contract}, lines~\ref{lst:msc_lines_start}--\ref{lst:msc_lines_end}, the contract verifies the authority key but omits an \verb|is_signer| check that confirms private key control for the authority.
Without the signer check, anyone can supply the correct key.
Second, attackers might craft accounts containing counterfeit state.
In this case, a missing owner check~(MOC) can trick the program into performing critical actions~\cite{slowmist_2024_solanasecurity,auditfirst_2022_solanavulnerabilities}.
\Cref{lst:exploit_contract}, lines~\ref{lst:moc_lines_start}--\ref{lst:moc_lines_end} read data of the \verb|vault_account| which is supplied as an argument to the instruction.
However, as shown in \Cref{fig:solana_account_model}, a program account owns the data account.
Without verifying that the data read originates from an account owned by the program, the program trusts unverified data.
Third, improper use of CPI allows malicious programs to gain unauthorized privileges~\cite{slowmist_2024_solanasecurity,auditfirst_2022_solanavulnerabilities}.
\Cref{lst:exploit_contract} lines~\ref{lst:cpi_lines_start}--\ref{lst:cpi_lines_end} invoke a CPI solely on an address given by the attacker.
As the accounts given to the CPI are all attacker-controlled, it is possible to manipulate or even write into those accounts and call arbitrary programs.
Since programs can delegate privileges to call targets, an attacker can call their own malicious program but with the caller's privileges.
That is, using a malicious arbitrary CPI~(ACPI) target, an attacker gains control over the program's PDA accounts.
In general, the design of Solana requires developers to add several specific checks even for simple use cases, e.g., a program requires an owner check once it reads the ID of an authority from its storage account.
The actual validation of the authority subsequently needs to check the public key and signer flag.
This makes errors and missing checks very likely.
Apart from missing checks and ACPI, Solana suffers from other security issues that we omit in our research.
Solana programs receive all accounts as untyped serialized bytes.
Hence, programs have to scrutinize the exact data structure and type.
Account confusion occurs when an attacker sends an account that is assumed to be the correct type due to common serialized forms but actually is of a different type, leading to unintended behavior.
Second, integer overflows and underflows occur when integers grow past their natural bit boundaries.
While Rust mitigates them in debug mode, the release mode optimizes overflows and wrapping behavior.
So, paths in bytecode that result in overflows may be pruned by the compiler resulting in many incorrectly detected or missed integer overflows.
Additionally, the Solana VM uses overflows explicitly to clear registers at runtime.
Thus, we ignore integer overflows to reasonably reduce false negatives.
Finally, reentrancy-like attacks reenter the original contract.
While Solana hinders Ethereum-like reentrancy, a versatile attacker could route through multiple contracts to reenter a call.
However, due to Solana's CPI depth limit of 4 calls, the pseudo reentrancy will only work once as direct recalls are not allowed.
Further, reentrancy attacks are only feasible in specific setups where the initial caller controls owner checked accounts of the victim as otherwise, writing into the data of the account is stopped by Solana.

\subsection{Differences to Ethereum}
\label{sec:diff-to-eth}
Unlike Ethereum, where trusted variables like \sym{msg.sender} guarantee transaction origin, Solana requires explicit signer and owner checks for each account, as programs can receive any account via instructions.
Third-party accounts and CPI targets are selected by index rather than by key, creating risks of account confusion and missing authorization checks that can lead to vulnerabilities.
Solana forbids the infamous Reentrancy vulnerability by limiting the depth of nested calls to 4 and disallowing direct reentrant calls.
While this technically still allows various Reentrancy attacks, these are mitigated in practice due to data ownership and decoupled contract logic.


\section{Related Work}
\label{sec:prev-oracles}
Prior research has explored smart contract security, yet few approaches address the unique challenges of Solana’s account-based, stateless execution.

\subsection{Vulnerability Detection in Ethereum}
\label{sec:rel-eth}
Ongoing smart contract security research largely targets Ethereum.
Static analysis approaches~\cite{Tikhomirov_2018_smartcheck, brent2018vandal, grech2019gigahorse, Bose2022sailfish, Tsankov2018securify} face scalability limits and often miss complex data/control-flow vulnerabilities, such as reentrancy~\cite{cecchetti2021}.
Dynamic techniques like fuzzing~\cite{Rodler_2023_efcf, ye2024, shou2023} are effective for diverse bugs, but face challenges on \ssc as described in \Cref{sec:rel-sol}.

Symbolic execution has gained traction for Ethereum due to small contract sizes~\cite{Luu_2016_making, mythril_git, Mossberg_2019_manticore, Torres2018osiris, Krupp2018teether, ethbmc2020, permenev2020verx, xue2020,Sunbeom_2021_smartest}.
Tools like Manticore~\cite{Mossberg_2019_manticore} and hybrid methods, which combine symbolic execution with fuzzing~\cite{mythril_git} or large-language models~\cite{Sunbeom_2021_smartest} provide comprehensive path exploration.
However, none of these approaches transfer to Solana due to its distinct programming model and vulnerabilities (see also \Cref{sec:diff-to-eth}).

Other lines of work include runtime monitoring~\cite{Torres2019aegis,sereum}, which requires platform modifications, and machine-learning-based analysis~\cite{sendner_2023_smarter,tx_based_classification,wu2024-advscanner}.
These methods show promise but rely on large datasets and reliable ground truth, which are currently unavailable for Solana.

\subsection{Vulnerability Detection in Solana}
\label{sec:rel-sol}
As Solana's popularity grows, ensuring its security is paramount.
Existing tools like VRust~\cite{Cui2022_nm} and \fds~\cite{smolka_2023_fuzz} cannot analyze all programs comprehensively.
VRust relies on Rust's MIR and requires source code, which is unavailable for \p{98} of deployed programs~\cite{smolka_2023_fuzz}, and raises many false positives---only 12 of 160 reported issues in eight projects were actual bugs~\cite{Cui2022_nm}.
Some bug oracles cover very specific cases~(e.g., unchecked Solana Program Library contract IDs), while others rely on small whitelists, raising completeness concerns.

\fds~\cite{smolka_2023_fuzz} integrates fuzzing into Solana's runtime and does not require source code, but cannot handle complex input structures of Anchor-based contracts used by \p{43} of deployed programs.
Solana's stateless nature makes fuzzing challenging: all input data---including multiple accounts and instruction fields---must be generated precisely to achieve meaningful coverage, otherwise, fuzzing risks low precision.
\fds ignores account trustworthiness in arbitrary CPI and requires extra knowledge of vulnerable functions for its missing key check~(MKC) oracle~\cite{smolka_2023_fuzz}.

Recent efforts apply formal verification to Solana~\cite{yuan2025}, but they target the underlying eBPF runtime and therefore face an entirely different class of vulnerabilities than Solana programs.
\citeauthor{andreina2024-userstudy} combined a developer study with static analysis of Solana bytecode~\cite{andreina2024-userstudy}. The analysis finds few bugs and only targets arbitrary program invocations, while the study shows developers often neglect security, highlighting the need for better vulnerability detection.

The bug detection mechanism is central to any vulnerability scanner, yet, existing tools' bug oracles vastly differ.
The bug oracles' limitations show that Solana vulnerabilities are complex and often undecidable, leading to large variations in the results.

\begin{figure*}[tb]
\centering
\includegraphics[width=1\linewidth, alt={Architectural overview of \tool: We lift sBPF and create symbols for accounts. Then, we symbolically analyze the program and check whether oracles are violated. Finally, a concrete report with a generated exploit is returned.}]{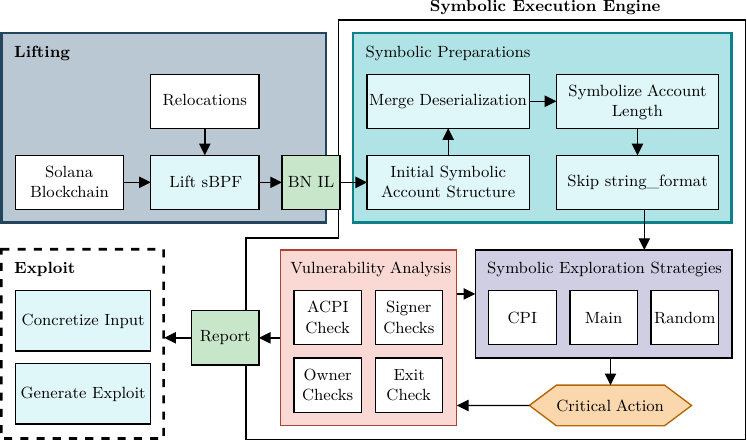}
\caption{Architecture of \tool}
\label{fig:arch}
\end{figure*}

\section{Design of \tool}
\label{sec:design}
This section presents the design of \tool, the first symbolic execution framework for effectively detecting diverse vulnerabilities in Solana contracts.
Our main goal is to avoid state explosion while at the same time achieving high coverage.
Further, we aim to cover the full set of contracts, meaning that we can analyze contracts for which source code is unavailable and contracts created with Anchor.
As a result, our approach tackles the key limitations of previous work in this field~\cite{Cui2022_nm,smolka_2023_fuzz}, and raises fewer false alarms.

At a high level, the functionality of \tool is separated into four phases, which are illustrated in \Cref{fig:arch}:
\begin{enumerate*}[]
\item \textbf{Lifting:} We dump the blockchain contracts' ELF files containing \sbpf bytecode instructions and prepare the files using \bn, i.e., lift the bytecode to IL and apply relocations;
\item \textbf{Symbolic Preparations:} We load the lifted bytecode into the symbolic execution engine and prepare the symbolic state optimized for \ssc;
\item \textbf{Symbolic Exploration:} Our symbolic exploration strategies advance the symbolic state towards critical actions, where attackers may influence the contract's state on the blockchain;
\item \textbf{Vulnerability Analysis:} Upon reaching points of critical actions, the symbolic state is evaluated to assess the vulnerability and a report is generated if missing checks are detected.
\end{enumerate*}
Last but not least, the detailed report facilitates automatic exploit generation, i.e., constructing proof-of-concept exploits.
\tool can automatically synthesize exploit transactions for reported vulnerabilities.
These exploits are valid for the analyzed path constraints, but they depend on a compatible initial blockchain state.
Hence, in practice, exploitability depends on whether such a state is reachable on-chain.
In future work, one could extend the capabilities of \tool to generate full precondition-establishing transaction chains, including interactions with additional contracts.
By focusing on exploit generation rather than transaction chains, \tool isolates and exposes security-critical behaviors with high precision.

\subsection{Initial Symbolic Account Structure}
Initially, we prepare the symbolic state and optimize the computationally expensive symbolic pointers and string operations as follows.

\paragraph{Deserialization}
\label{sec:dese}
All Solana contracts deserialize account data from the input, posing two challenges for symbolic execution.
First, deserialization produces many similar states, e.g., an account's signer field is either true or false.
Just from the three boolean account fields (signer, writable, executable), $2^3=8$ states are created.
For example 10~accounts result in $(2^3)^{10}\approx {10}^9$ states.
We address this by merging similar states into a single state during deserialization (\Cref{sec:impl-dese}), leveraging common deserialization routines from the Solana framework.

Second, the account data structure contains fields of variable length, in particular its main \sym{data} field, which causes the deserialized state to contain variably \emph{positioned} buffers.
The serialized data stores accounts sequentially, making each account's start depend on preceding lengths.
The symbolic pointers to the account data would accumulate these lengths, e.g., the pointer for the tenth account will be the sum of all nine preceding buffer lengths plus a constant, which would exceed the solving capabilities of the symbolic engine.
We avoid variably positioned buffers by using a constant maximum size during deserialization and symbolizing the lengths afterward in the deserialized state (cf.~\Cref{sec:impl-hyrid-length}).

\paragraph{String Formatting}
String operations are a challenge for symbolic execution, in particular, formatting numbers.
The bytecode to convert a symbolic number to its decimal string representation produces complex dependencies between the symbolic bits and the resulting string.
Consequently, when this string is used, the solver has to solve for a string representation that satisfies the constraints of the symbolic number, which often stalls the execution engine.
However, \ssc almost exclusively use string formatting for logging, which does not affect the outcome of transactions.
Therefore, we search for string format/logging calls and skip these functions.

Here, our starting point is the Solana system call used for logging and the string format function of Rust, which is very stable and can be identified from its bytecode.
Then, we analyze the calls to these functions and identify sequences of \func{format}, \func{log}, and string deallocation to ensure that the formatted string is only used for logging.
When all checks succeed, we skip this call to \func{format} during symbolic execution.

\subsection{Exploration for Critical Actions}
During the main analysis stage, we use several exploration strategies to prioritize execution paths in the contracts that are more likely to be security-critical, which decreases the time for bug discovery.
That said, given sufficient time, \tool will still analyze the complete contract.
In general, our exploration strategies prefer uncovered paths and prune paths that deterministically lead to \func{abort} (cf.~\Cref{sec:impl-pruning}).
Further, switching strategies and selecting different execution paths prevents stalling of the symbolic exploration in complex states.
For \tool, we developed three strategies to target critical actions within \ssc.

\begin{enumerate}[(1),nosep,wide]
\item The \emph{CPI} exploration only follows paths that are statically determined to reach the Solana system call for \emph{CPI}.
\item The \emph{Main} exploration identifies the core switch tree, which \ssc commonly employ to dispatch the desired high-level instruction number to its bytecode, and leads the symbolic execution into the leaves of this tree.
This is where most contracts implement their logic.
\item The \emph{Random} exploration extends coverage beyond common code paths, selecting paths at random.
This is especially useful to guide the symbolic executor through code sections that the other strategies might not consider interesting.
\end{enumerate}

In our evaluation, we leverage all explorations and apply each strategy uniformly to uncover diverse bugs.
Specifically, we execute each of the three exploration strategies round-robin until a critical action is found, or at most for \SI{10}{\minute}.
This time limit has proven sufficient for progress while remaining small enough to avoid stalling in undecidable path constraints.
Further, a strategy is skipped if its target cannot be reached (e.g., a syscall for CPI).

\subsection{Vulnerability Analysis}
\label{sec:vuln-analysis}
The symbolic exploration of \tool analyzes the control flow to find critical actions that are vulnerable without proper validation.
When such a state is reached, we test for the conditions described in this section and generate a report if validation is missing.
\tool reacts to unintended program behavior that results in security-critical actions, regardless of whether the root cause is a missing validation, flawed access-control logic, incorrect state transitions, or subtle memory-safety violations.
The underlying symbolic execution reasons about full program behavior and can still expose traditional low-level root causes (e.g., memory corruption effects) when they manifest as vulnerable Solana critical actions.

We begin by outlining the types of checks we assess, followed by criteria for identifying when a missing check constitutes a vulnerability in the context of certain critical actions.

\paragraph{Signer Checks}
Every input account may be a signer, indicated by the \sym{signer} field.
\tool checks which signer flags are required by the path constraints for each critical action.
A program function without any signer check can be called by anyone, which is usually not desired (cf.~\Cref{sec:oracle-eval}), so we consider this a missing signer check bug.
Since full business logic is unavailable (without source code), it is not possible to precisely identify which accounts require signer checks.
Thus, we consider a single signer check on any account to be sufficient.

Since paths may branch on the signer field without affecting the bottom line (similar to deserialization, cf.~\Cref{sec:dese}), we double-check and re-execute the path to the current state with the same constraints as before, but all signer fields set to false.
Thus, we can confidently evaluate if any signer fields are checked on the path to the critical action, and if not, report a \emph{Missing Signer Check~(MSC)}.

\paragraph{Owner and Key Checks}
There are multiple ways to verify the authenticity of an account.
\begin{enumerate*}
\item
Every account has a unique address/ID, bound to its key pair.
This can be used to identify an account uniquely.
\item
Accounts have an owner, either a contract, or the system program in the case of user accounts.
\ssc can verify the ownership of accounts, in particular, that the contract itself owns an account.
\item
Programs may cryptographically derive related accounts (PDAs) using one account's public key.
Since an attacker cannot control the address of the PDA, this account is also considered trustworthy.
\item
An account that is written to is considered trustworthy because accounts can only be modified by the owner.
As a result, changing account data implies an owner check.
This also holds if an account write occurs after the critical action, as the runtime discards any changes when encountering an exception.
Consequently, the accounts vulnerable to missing owner checks are only read.
\end{enumerate*}

In the first two cases, the owner or key field is compared to other data.
Since the bytecode does not document which value a key/owner field must be compared with, we consider any check on the key/owner valid.
The fourth case, implicit owner checks, requires the execution of the contract from the entry point until the contract gracefully terminates.
Hence, we gather account write operations not only until we reach a critical action but also continue to collect them afterward.

\paragraph{Authority Checks}
Contracts frequently implement authority verification by combining owner checks and signer checks, i.e., they validate that a \emph{trusted} account has \emph{signed} the transaction.
In this context, we consider an account trusted if the contract has validated its public key against either a constant value or the data stored in an owner-checked account.
We consider this authority check sufficient to protect a transaction from manipulated data.
Assessing more complex scenarios, such as multi-authority schemes, requires additional information about the purpose of each account, which is not available at the bytecode level.

\subsection{Vulnerabilities}
Vulnerabilities arise when critical actions are performed without adequate input validation.
In the following, we describe critical actions and how we validate the checks to identify \emph{vulnerable critical actions}.
We focus on generic oracles that can be applied to a wide range of contracts.
In particular, we exclude oracles for vulnerabilities that are already fully mitigated by current toolchains:
The safe-math option of Rust generates code safe from integer overflows, type confusion is averted by serialization libraries that use type IDs, and the updated Solana SDK mitigates the missing system account check known from the Wormhole incident.

\paragraph{Unchecked Account Writes}
Modifying persistent data stored on the blockchain, i.e., writes to account data, is a critical action.
This includes metadata fields like the account's lamport balance as well as contract-specific data stored in data accounts, such as token balances.

We consider account writes vulnerable if they lack sufficient owner/key checks~(MOC) or are performed without a signer check~(MSC).
In addition, the execution path must exit gracefully; otherwise, the runtime will rollback and invalidate the transaction.
We consider any account checked for which any of the following is true:
\begin{inparaenum}[(1)]
    \item Owner field compared
    \item Public key compared to constant value
    \item Data field written or changed
    \item PDA derived from public key.
\end{inparaenum}

If there is any used but unchecked account ($\var{read\_accounts} \setminus \var{checked\_accounts}$), and there is no trusted signer (\emph{authority check}), we classify this as a missing owner check vulnerability.
Both checks may also be missing in combination (MOC/MSC), which grants even more freedom to an attacker.

\paragraph{\ACPI (ACPI)}
A CPI is a (privileged) delegated call of a Solana program to other contracts.
This is a critical action because it allows the target to perform state modifications on behalf of the caller.
\tool tests each CPI to determine if the target is arbitrary, whether the source of the target address is not owner-checked~(MOC), and if there is no signer check~(MSC).
If all these conditions are met, we consider this a vulnerability.
If either an MOC or an MSC is present, the CPI may still be vulnerable.
However, analyzing such cases requires access to the source code and a deep understanding of the contract's business logic.


\section{Implementation}
Implementing \tool required substantial engineering efforts.
First, we added bytecode lifting (\SI{400}{loc}) to the reverse-engineering toolkit \bn~\cite{binaryninja,bn-ebpf-solana} including precise instruction semantics and calling conventions for automated symbolic execution.
Next, we built symbolic execution for Solana atop the SENinja~\cite{seninja} AMD64 framework, requiring \SI3{kloc}, forming the foundation for implementing Solana-specific oracles and vulnerability detection (\SI4{kloc}).
The implementation follows the design in \Cref{sec:design} to search critical actions using custom symbolic exploration and vulnerability oracles.

We also optimized the symbolic exploration to handle variable data lengths (\Cref{sec:impl-hyrid-length}) and string formatting (\Cref{apx:strings}).
For example, the Neodyme workshop examples (\Cref{sec:neodyme}) print formatted log strings, which caused a path explosion and never reached the critical actions.
Further, Anchor contracts perform input-length checks; fixed lengths fail the check and fully symbolic lengths cause a constraint explosion.
The hybrid symbolic length avoids both issues (\Cref{sec:impl-hyrid-length}).

\subsection{Exploration and Static Path Pruning}
\label{sec:impl-pruning}

\paragraph{Exploration}
\tool begins exploration at the entry point, executing instructions until the next symbolic branch.
This is when the subsequent instruction is ambiguous, e.g., due to a conditional jump that depends on a symbolic value, which causes a state fork into two or more program states to execute.
Similar states may be merged (cf.~\Cref{sec:impl-dese}), but generally, the executor must select one state to follow.
Effective selection or pruning of states can greatly improve symbolic exploration efficiency.
\tool keeps track of covered instructions and selects the state to execute as follows:
\begin{enumerate*}
\item Continue executing the current state if its instructions are not yet covered.
\item Select any uncovered deferred state.
\item Continue executing the current state if it is not at the program exit or in an unsatisfiable condition.
\item Select the most recently forked state.
\item Select a state at random.
\end{enumerate*}

\paragraph{Path Pruning}
The exploration strategies \emph{CPI} and \emph{Main} select specific addresses in the contract executable, which are more likely to contain vulnerabilities.
During exploration, the \emph{active} states are evaluated to determine whether the targets are still reachable using the control-flow and caller graphs.

First, a state may dominate and directly reach a target.
Second, a state may be within a caller of a target, hence, the target is reachable if the call to the target function is reachable.
Third, a parent function in the current call stack may be within the caller graph, hence, the target is reachable if the current state can return.
Otherwise, a state is rejected and not followed by the execution strategy.

Notably, this avoids common \sym{panic!} error handling paths that merely print a message and \sym{abort} the execution.
These paths cannot reach the targets nor return to a function that can.

\subsection{State Merging after Deserialization}
\label{sec:impl-dese}
Symbolic state merging combines different states' path constraints into a single state with memory constraints~\cite{angr}.
\begin{figure}[tb]
\centering
\begin{lstlisting}[style=rust-style,basicstyle=\small\ttfamily,
xleftmargin=0pt,
framesep=0pt,numbers=none,
caption={The listing shows the decompiled bytecode of the signer field's deserialization.},
label={lst:signer_field_deserde}
]
    for i in 0..number_of_accounts
        account[i].is_signer = 1
        if read_serialized_signer() == 0
            account[i].is_signer = 0
\end{lstlisting}
\end{figure}

Executing \Cref{lst:signer_field_deserde}, symbolic execution creates two states, one where \verb+is_signer+ is~1 and the condition was assumed false, and the other where the \verb+if+ branch was taken and \verb+is_signer+ is~0.
State merging combines these two states into one conditional memory expression, i.e., \verb+is_signer+ is \verb+read_serialized_signer() == 0 ? 0 : 1+.
This eliminates many redundant states, e.g., deserializing three boolean fields for 10~input accounts, reduces the redundancy by $(2^3)^{10}\approx {10}^9$, significantly easing main analysis.

To optimize state merging, we identify the point in the control flow where states differ minimally, as with the signer field in the example above.
We detect the loop deserializing accounts and merge states after every loop.
Our algorithm analyzes the entry function and its callees for the characteristic comparison (\verb+jeq/jne r*, 0xff+) of the constant first byte in the account structure.
Then, it traverses the control-flow graph to find the enclosing loop that iterates all accounts.
Lastly, we find the exit condition in the basic blocks of the loop, which will be the point for state merging.
Usually, this yields a single deserialization loop, and otherwise, state merging is disabled.

\subsection{Hybrid Symbolic Data Length}
\label{sec:impl-hyrid-length}
\begin{figure}[tbh]
    \centering
    \includegraphics[alt={We deserialize accounts and symbolize their data size to analyze all accounts.}]{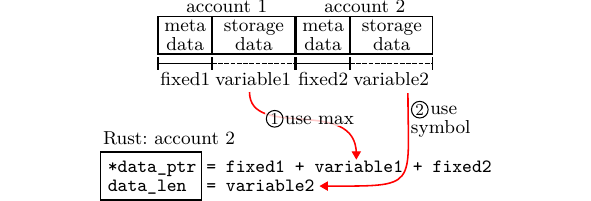}
    \caption{The position of the second account is symbolized with the fixed offset of all preceding fixed metadata sizes and each account's maximum storage data size.}
    \label{fig:hybrid_length}
\end{figure}

The serialized transaction context, containing all account fields, is parsed into Rust objects.
To avoid symbolic pointers---which produce complex constraints that overload the solver---we limit account data lengths during deserialization~\circleone and release these constraints afterward~\circletwo (\Cref{fig:hybrid_length}).
Symbolic variables represent data lengths, constrained to the maximum supported length.
Deserialization creates Rust strings with fixed pointers, and once all accounts are deserialized, the length constraints are removed to allow variable data sizes.

\subsection{String Format Detection}
\label{apx:strings}
String operations are a challenge for symbolic execution, in particular, formatting numbers.
The bytecode to convert a symbolic number to its decimal string representation produces complex dependencies between the symbolic bits and the resulting string.
Consequently, when this string is used, the solver has to solve for a string representation that satisfies the constraints of the symbolic number, which often stalls the execution engine.
However, \ssc almost exclusively use string formatting for logging, which does not affect the outcome of transactions.
Therefore, we search for string format/logging calls and skip these functions.

Here, our starting point is the Solana system call used for logging and the string format function of Rust, which is very stable and can be identified from its bytecode.
Then, we analyze the calls to these functions and identify sequences of \func{format}, \func{log}, and string deallocation to ensure that the formatted string is only used for logging.
When all checks succeed, we skip this call to \func{format} during symbolic execution.


\section{Evaluation}
\label{sec:evaluation}

We evaluate the correctness of \tool using open-source contracts and prior datasets with known vulnerabilities.

\subsection{Vulnerability Oracles}
\label{eval:oracles}
Because source code is unavailable for most public contracts, large-scale measurements alone are insufficient to validate findings against developer intent.
Therefore, we establish a baseline for \tools oracles by analyzing known vulnerabilities from the Neodyme workshop and from the datasets of VRust~\cite{Cui2022_nm} and \fds~\cite{smolka_2023_fuzz}.
Overall, the evaluation shows that \tool detects known true positives while reporting substantially fewer false positives.

\paragraph{Neodyme}
\label{sec:neodyme}
Each Neodyme Workshop contract demonstrates one specific vulnerability.
Level~0 implements a wallet service that stores user lamports in contract-managed wallets.
The contract contains a missing owner check, allowing attackers to steal funds from other users.
During withdrawals, four entities must be verified:
The \textbf{Vault} holds all user assets and is identified from wallet data.
When lamports decrease, the runtime implicitly enforces deduction permissions.
The \textbf{Authority} owns the funds and must authorize the transaction via a signature check.
The \textbf{Payout destination} receives lamports and is not further validated because payout is already authorized by the user/owner.
The \textbf{Wallet} links these entities and stores user ID, balance, and vault ID.
However, because wallet ownership is not verified, attackers can construct counterfeit wallets for arbitrary vault owners and withdraw all managed funds.
\tool detects that wallet account data is checked, but wallet owner field is not, and reports this vulnerability.

Level~1 demonstrates a missing signer check: wallet ownership is verified, but the authority signature is not, allowing unauthorized withdrawals from existing wallets.
\tool detects this as a (lamport) state modification without a signer check.

Level~2 and~3 address integer overflows and type confusion, which are mitigated by compiler-generated safe math and type identifiers.
Consequently, we did not implement corresponding oracles and \tool did not detect an error.

Level~4 contains an \acpi vulnerability: attackers directly control the CPI target and can inject arbitrary logic into the control flow.
\tool reports this when the call target is neither constant-checked nor read from an owner-checked account.

\paragraph{VRust}
The static source-code analysis tool VRust reports results for eight projects.
According to the authors, VRust found 30~integer overflows (21~FP), 3~key check errors (2~FP), and 2~vulnerable CPIs.
We analyze the same projects and versions with \tool.

First, we consider integer overflows out of scope for bytecode analysis (cf.~\Cref{sec:prev-oracles,sec:design}).
VRust's own results indicate that integer-overflow detection is unreliable; even with source code, the false-positive rate is \p{70} (\p{92} before manual triage).
Second, VRust reports "one wrong key check in spl-governance"~\cite{Cui2022_nm}, but the raw report lists 11 missing key checks with two patterns.
In the first case, the owner check is done after the potentially unsafe read, which is fine in Solana.
In the other case, result data is written back to an account, thus it is implicitly owner-checked by the runtime.
Hence, analysis tools must account for both post-action checks and implicit runtime checks.
Third, \tool attests that all system CPIs are correctly validated when using the updated Solana program library.

\tool further reports 11~missing signer checks and one \acpi in spl-instruction-padding~\cite{spl}.
Because that program intentionally wraps arbitrary invocations, this behavior is expected.
Additionally, spl-governance~\cite{spl} allows unsigned version-number updates.
While this is not directly exploitable on its own, it can be used in downgrade or type-confusion attacks.
The transfer-lamports example in the Solana program library~\cite{spl} is intentionally minimal and contains no checks, leaving it vulnerable and contract-owned accounts drainable by attackers.

\begin{table}[tb]
    \centering
    \begin{subtable}[t]{0.45\textwidth}
        \caption{Comparison of \tool to \fds}
\label{tab:fuzzdelsol}
\centering
\begingroup
\newcommand{\cf}{{\color{orange}F}\xspace}
\newcommand{\cs}{{\color{cyan}S}\xspace}
\newcommand{\yf}{{\color{orange}\yes}}
\newcommand{\nf}{{\color{orange}\no}}
\newcommand{\ys}{{\color{cyan}\yes}}
\begin{adjustbox}{max width=\linewidth}
\begin{tabular}{ll|cccccccccccccc}
\toprule
Type &      & \rotatebox{90}{\shortcon{3od3X7QN84FTonkyXbQiT1ydxcT9P1jBcA9mbgD3jnsW}} & \rotatebox{90}{\shortcon{3VtjHnDuDD1QreJiYNziDsdkeALMT6b2F9j3AXdL4q8v}} & \rotatebox{90}{\shortcon{3w57iMhv5Zk5VDuTe5dspm2FzE9zQhscCb9CZpAKobPW}} & \rotatebox{90}{\shortcon{4hPkNV2WsgPW1wHHcQebvV7GLyLgdDDLEx3Pu6LzJP5N}} & \rotatebox{90}{\shortcon{4M2fancicHbUtMLcMNmbi97YngFoqBcnFk5D31JjjStx}} & \rotatebox{90}{\shortcon{6LanqAFCbucXWSG35ssij4kFDTWJ25BY7d6hbR2szqi}} & \rotatebox{90}{\shortcon{7FWEcVG1YRW7evGR3bXgu47ge8m6Je7BQuvTzMbn9p7p}} & \rotatebox{90}{\shortcon{9a5dihgNgBhWnjmRDJ8rUy4ihetvgMmjaPk7NGdsjZP9}} & \rotatebox{90}{\shortcon{9tSWsKwtDL6YseLuh1haGFJk312uu9HGyrnVa5XH11yy}} & \rotatebox{90}{\shortcon{9WoLnfjLKk1EBtkABhe3vcA8CLogsbs3XBoddn8h849B}} & \rotatebox{90}{\shortcon{GQ6qchUsofiK7rzeFg5jbvpHcJ7pNnfL4yfwaYrB1u6K}} & \rotatebox{90}{\shortcon{H5rpfCD6hLFCPCfxxqjGg94Gqoigqfk7afhqGLu1nPSG}} \\ 
\midrule

\multirow{2}{4em}{ACPI} & \cf  & \yf &     & \yf & \yf &     &     & \yf &     &     & \yf &     &     \\ 
 & \cs  & \ys &     & \ys & \ys &     &     & \ys &     &     & \ys &     &     \\ 
 \midrule
\multirow{2}{4em}{MSC} & \cf   &     &     &     &     & \yf & \yf &     &     & \yf &     &     &     \\ 
 & \cs   &     &     &     &     & \ys & \ys &     & \ys & \ys &     & \ys &     \\ 
 \midrule
IO & \cf    &     & \yf &     &     & \yf &     &     & \nf & \yf &     & \yf & \yf \\ 
\midrule
\multirow{2}{4em}{MOC} & \cf   &     &     &     &     &     &     &     &     &     &     &     &     \\ 
 & \cs   &     & \ys & \ys &     & \ys & \ys &     &     & \ys &     &     & \ys \\ 
\bottomrule
\end{tabular}
\end{adjustbox}

\medskip
\hfill \cf: \fds\qquad \cs: \tool\hfill\null
\endgroup
\end{subtable}%
\hfill
\begin{minipage}[t]{0.5\textwidth}
\begin{subtable}[t]{\textwidth}
    \caption{Anchor Contract Transparency Program}
    \label{tab:anchor-apr}
    \begin{adjustbox}{max width=\linewidth}
    \begin{tabular}{lcll}
        \toprule
        Project & Anchor & \tool & Manual Analysis \\
        \midrule
        quarry-registry & \yes & MSC & Deterministic data \\
        serum-dex & \no & MSC & Zeroed account \\
        sol-did & \yes & MOC & Pot. Exploitable \\
        spl-token-2022 & \no & MSC & Fresh account \\
        \bottomrule
    \end{tabular}
    \end{adjustbox}
\end{subtable}%
\bigskip

\begin{subtable}[t]{\textwidth}
    \caption{Large-scale Analysis of \tool}
    \label{tab:large-scale}
    \begin{adjustbox}{max width=\linewidth}
    \centering
    \begin{tabular}{lr}
    \toprule
    & Contracts \\
    \midrule
    Total Analyzed & \totalanalyzed \\
    \midrule
    Unchecked Account Write & 374 \\ 
    -- Missing Owner Check (MOC) & 117 \\ 
    -- Missing Signer Check (MSC) & 241 \\ 
    -- MOC/MSC & 33 \\ 
    \midrule
    \ACPI & 100 \\
    \midrule
    Total Reported & \totalbugs \\
    \bottomrule
    \end{tabular}
    \end{adjustbox}
\end{subtable}
\end{minipage}
\medskip
\caption{The tables show the evaluation results of \tool.}
\end{table}

\paragraph{\fds}
\fds is the state-of-the-art fuzzer for \ssc.
We analyze the same contracts reported by \fds~\cite{smolka_2023_fuzz}.
As shown in \Cref{tab:fuzzdelsol}, \tool classifies these contracts as vulnerable and largely matches the reported bug types.
Note that \fds's results are based on a large-scale study of all contracts on the blockchain
with a timeout of 5 minutes per contract.
In comparison, \tools large-scale study identified \totalbugs~contracts with missing checks (cf.~\Cref{tab:large-scale}).
Since both analyses are greybox techniques, source code was unavailable for manual ground-truth verification.

\paragraph{Anchor Contract Transparency Program}
The Anchor project provided a registry of verified Solana contracts from April to November 2022, containing 666 builds from about 120 projects.
\p{90}~of the projects in this registry are Anchor-based.
Because Anchor provides strong built-in mitigations, we use this registry as a baseline to assess the false-positive rate of \tool.
Only four vulnerabilities were reported across 120~projects (\Cref{tab:anchor-apr}).
The most notable report concerns the Anchor-based project \mbox{\emph{sol-did}}~\cite{sol-did}, which manages decentralized identities in \sym{DidAccounts}.

Solana requires accounts to maintain a minimum \sym{rent}-exempt balance determined by account size.
The function \sym{resize} in sol-did allows users to grow or shrink their \sym{DidAccounts} as needed.
Because resizing changes required \sym{rent}, Anchor adjusts balances automatically and pays out surplus \sym{rent}.
However, \tool found that the resized \sym{DidAccount} is not verified, allowing arbitrary resizing.
Our manual analysis confirms this observation.
In sol-did, the secp256k1 signature used for Ethereum authority checks is constructed solely from the requested \sym{DidAccount} size.
Hence, an attacker can impersonate an Ethereum authority and resize \sym{DidAccounts}.

Enlarging foreign \sym{DidAccounts} is unattractive because the attacker must pay additional \sym{rent}.
Anchor also prevents zeroing an account by enforcing that account metadata still fits.
Yet attackers can shrink accounts to the minimum size and collect surplus above the lower \sym{rent}-exemption threshold, including by racing transactions after a user increases account size.

Manual analysis of the remaining contracts shows that these functions are indeed unchecked and correctly reported by the oracle, but their practical impact is often constrained by business logic or deployment context.
We discuss such challenges that arise in automatically examining logic-related vulnerabilities in more detail in \Cref{sec:oracle-eval}.

\subsection{Large-scale Analysis}
\label{sec:eval:large-scale}
We analyze \totalanalyzed~contracts deployed on the Solana blockchain of which \num{3763}~(\p{43.2}) are Anchor-based.
To the best of our knowledge, this is the largest vulnerability study in Solana as of now.
We run \tool on two machines with \gb{240}~RAM each, processing contracts in groups of 8~parallel jobs.
We limit the time per job to \SI2h to keep the overall analysis time feasible.
Our results in \Cref{tab:large-scale} suggest that a notable share, about \p{5.4}, of the deployed contracts contain at least one vulnerability.
Note that \Cref{tab:large-scale} counts contracts, so categories overlap.
Since most of these contracts are deployed anonymously, we examined a set of 11~contracts in detail pertaining to 6~open-source projects (cf.~\Cref{sec:oracle-eval}).

To further assess oracle validity, we manually analyze a random sample of \num{30}~anonymous contracts.
We exclude contracts with just the deployment transaction, so we can use information from the transaction history to infer the use cases of the contracts.
Since source code is unavailable, we combine multiple signals to infer contract purpose: \tools report, similar open-source code, bytecode inspection, and historical transaction data.
Our results, which are described in detail in the artifact, show that for the 30~contracts, we discovered \p{87} true bugs.
We note that several contracts currently do not own or manage targetable accounts or lamports.
For example, some contracts contain an ACPI vulnerability that allows attackers to modify private accounts.
However, since the contracts do not rely on these accounts, they exhibit only benign behavior---even in the presence of the bug.
Similarly, we observed contracts with technically valid draining paths where no relevant contract-owned assets were present in practice.
During this evaluation, we did not observe false positives caused by incorrect oracle predicates.
However, some reports describe vulnerabilities that are technically present but context-dependent in impact.
For example, a contract may expose a path that can drain contract-owned accounts, while the observed deployment does not hold targetable assets.
We further observe 14~missing signer checks, i.e., attackers can impersonate legitimate users and perform restricted actions.

\begin{figure}[t]
\centering
\includegraphics[width=0.9\linewidth, alt={We show the coverage achieved by \tool over the analyzed time. We show that coverage increases for longer analysis times but decrease over the size of the contract.}]{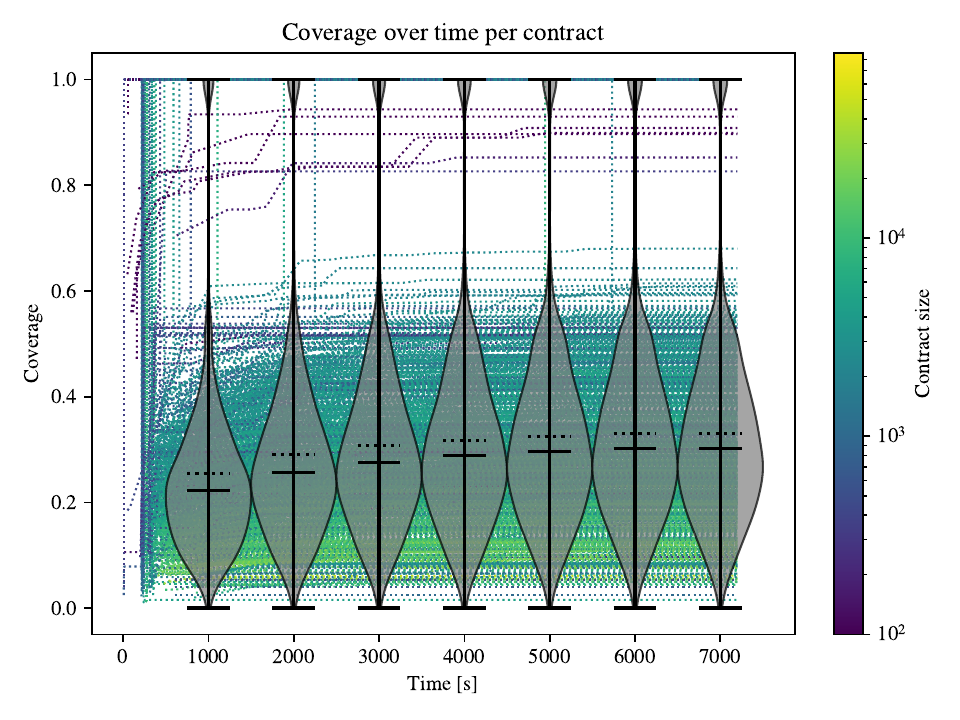}
\caption{Coverage ratio over time per contract. Violins summarize the individual runs at each point in time. Color encodes total contract size in basic blocks.}
\label{fig:coverage}
\end{figure}

\subsection{Coverage}
\label{sec:eval-coverage}
\Cref{fig:coverage} shows the coverage during our large-scale analysis, which follows the typical progression of binary analysis.
Most basic blocks are covered early, then coverage grows incrementally until reaching a plateau.
Upon termination of the analysis, \tool reached a mean coverage of \p{33} ($\sigma=\p{18.5}$).
For binary-only analysis, coverage percentages must be interpreted carefully because bytecode contains unreachable code, e.g., unused functions from linked libraries such as string-formatting routines.
Further, \tool excludes paths for error handling that abort the transaction (cf.~\Cref{sec:impl-pruning}).
An analysis of 239 contracts, where \tool covered all relevant paths, indicates a mean unreachable code ratio of \p{84.6} ($\sigma=\p{18.6}$).


\section{Case Studies and Lessons Learned for Solana Vulnerability Analysis}
\label{sec:oracle-eval}
This section presents three case studies and discusses their root causes, security impact, and implications for automated analysis.
The first two case studies present previously unknown vulnerabilities detected by \tool.
In the third case study, Wormhole, we discuss limitations of automated oracles.

\subsection{Case Study: Censo Solana Wallet}
The Censo~\cite{censo} Solana Wallet program (formerly Strike Wallet) provides multi-signature wallets through an $m$-of-$n$ signature scheme.
Our analysis reports a missing signer check in the Censo Solana Wallet program, whose source code was available until recently.
The program is written against Solana's plain Rust SDK, i.e., it does not use Anchor.
Through source code analysis, we found two vulnerabilities caused by this missing signer check.
First, wallets may be terminated without authorization, which still pays out to the rightful owner.
Because the attacker cannot directly profit, this is a griefing attack.
Second, the wallet upgrade function allows changing the recipient of the termination payout.
Combined, the second bug escalates the griefing attack into a profitable attack.


\begin{figure}[tb]
\begin{lstlisting}[escapechar=@, style=rust-style, caption={The code shows Censo's cleanup handler in Rust. While the program performs several sanity checks, the transaction signer is never checked. We highlight a possible fix in green.}, label={lst:censo-cleanup}]
pub fn handle(program_id: &Pubkey, accounts: &[AccountInfo]) -> ProgramResult {
  let accounts_iter = &mut accounts.iter();
  let wai = next_wallet_account_info(accounts_iter, program_id)?;
  let cai = next_program_account_info(accounts_iter, program_id)?;
  let rrai = next_account_info(accounts_iter)?;
@\settikzmark{hl1s}\greenplus\,\,@assert!(rrai.is_signer);@\label{censopatch}\settikzmark{hl1e}@
@$\cdots$@
   collect_remaining_balance(cai, rrai)?;
   Ok(())
}
\end{lstlisting}
\begin{tikzpicture}[remember picture, overlay]
\begin{scope}[on background layer]
  \diffhighlightadd{hl1s}{hl1e}
\end{scope}
\end{tikzpicture}
\end{figure}


\paragraph{Unauthorized Wallet Termination}
When a multi-signature wallet data account is created, one authority pays the rent and stores its address for reimbursement when the wallet is deleted.
Censo wallet's cleanup handler (cf. \Cref{lst:censo-cleanup}) requires three accounts:
\begin{enumerate*}
    \item the multi-signature wallet,
    \item the account to be deleted, and
    \item the rent return account, i.e., the authority's wallet account.
\end{enumerate*}
Ideally, only the wallet authority should call this handler, but the missing signer check allows anyone to delete accounts.
Since the rent return account is validated, the attacker receives no direct profit (griefing attack).


\begin{figure}[tb]
\begin{lstlisting}[escapechar=@, style=rust-style, caption={The code shows Censo's migration handler in Rust. We highlight a possible fix in green.}, label={lst:censo-migration}]
fn migrations() -> @\label{migstart}@BTreeMap<u32, MigrationFunction> {
    BTreeMap::from([
        (MIGRATION_TEST_VERSION, migration_test as MigrationFunction),
        (2, v2_to_v3_add_name_to_signers as MigrationFunction),
        (3, v3_to_v4_add_latest_activity_at as MigrationFunction),
    ])
@\label{migend}@}

pub fn handle(program_id: &Pubkey, accounts: &[AccountInfo]) -> ProgramResult {
    let accounts_iter = &mut accounts.iter();
    let sai = next_program_account_info(accounts_iter, program_id)?;
    let dai = next_program_account_info(accounts_iter, program_id)?;
    let rrai = next_account_info(accounts_iter)?;
@\settikzmark{hl2s}\greenplus~@   assert!(rrai.is_signer);@\label{censopatch2}\settikzmark{hl2e}@
@$\cdots$@
    if let Some(migrator) = migrations().get(&source_version) {
        migrator(
            &sai.data.borrow(),
            &mut dai.data.borrow_mut(),
            rrai.key,);
        Ok(())
@$\cdots$@}
\end{lstlisting}
\begin{tikzpicture}[remember picture, overlay]
\begin{scope}[on background layer]
  \diffhighlightadd{hl2s}{hl2e}
\end{scope}
\end{tikzpicture}
\end{figure}

\paragraph{Unauthorized Migration}
The second vulnerability is an unauthorized migration, shown in \Cref{lst:censo-migration}.
When the contract data structure changes, the Censo Wallet migration handler updates all accounts.
We discovered that the migration handler uses the caller's public key as the rent return account instead of preserving the original value.
Because it also lacks a signer check, an attacker can migrate accounts and insert their own address as rent return account.
In combination with the unauthorized wallet termination bug, this allows attackers to profit from un-migrated wallet accounts.
Note that both the migration from version~2 to~3 and the migration from version~3 to~4 are affected, as the migration functions do not include any checks.

\subsection{Case Study: Elusiv}
Elusiv~\cite{elusiv,elusiv_git} enables private token transfers via zero-knowledge proofs and multiparty computation~\cite{elusiv_quicknode,elusiv_eiger}.
Elusiv consists of the core protocol, the Warden, and the ZEUS protocol.
Users deposit tokens in the core token pool, where the funds are hidden among other tokens and tracked via commitments.
Receivers withdraw funds from the pool using a zero-knowledge proof.
The ZEUS protocol verifies benign users to block malicious actors, while the Warden processes requests and prevents flagged users from interacting with the core protocol.

\begin{figure}[tb]
\begin{lstlisting}[style=rust-style,caption={The code shows Elusiv's vulnerable function in Rust. While the program checks that the hashing account is active, the transaction signer is never checked.},label={lst:elusiv-bug}]
pub fn compute_base_commitment_hash(
  hashing_account: &mut BaseCommitmentHashingAccount,
  _hash_account_index: u32,) -> ProgramResult {
  guard!(hashing_account.get_is_active(),
      ElusivError::ComputationIsNotYetStarted);
  compute_base_commitment_hash_partial( hashing_account)
}
\end{lstlisting}
\end{figure}

\Cref{lst:elusiv-bug} shows the vulnerable contract code.
\tool reports a missing signer check in Elusiv's \sym{compute\_base\_commitment\_hash}.
This function initializes hashing for a \sym{BaseCommitment\allowbreak{}Hashing\-Account}, which is a crucial step in tracking user funds.
Because the function does not verify the signer for the tracked funds, any valid account suffices; an attacker who pays the transaction fee can start hashing any commitment.

Normally, contracts process only user-authorized accounts.
Here, the attacker triggers a deterministic calculation on trusted data using their own funds, so we classify this bug as low risk.
Even with high transaction volume, it is unlikely to affect Elusiv's performance or expose sensitive commitment data.

At bytecode level, the oracle correctly flags an unauthorized modification of foreign data.
Yet, the contract logic renders this bug harmless.
This case study demonstrates that evaluating real user impact still requires a deep understanding of contract logic.


\begin{figure}[tb]
\begin{lstlisting}[style=rust-style,caption={The listing shows the vulnerability that led to Wormhole accepting an arbitrary secp256k account.},label={lst:wormhole}]
    // The previous ix must be a secp verification instruction
    let secp_ix_index = (current_instruction - 1) as u8;
    let secp_ix = solana_program::sysvar ::instructions::load_instruction_at(
        secp_ix_index as usize,
        &accs.instruction_acc.try_borrow_mut_data()?,
    ).map_err(|_| ProgramError::InvalidAccountData)?;
\end{lstlisting}
\end{figure}

\subsection{Case Study: Wormhole}

Wormhole is a cross-chain bridge that transfers assets between blockchains such as Solana and Ethereum.
In 2022, Wormhole was exploited through a missing key check in \sym{load\_instruction\_at}.
This allowed an attacker to supply a counterfeit system account and forge the payment-signature context, creating Solana tokens without corresponding payments.
\Cref{lst:wormhole} shows the vulnerable function.

Because contracts for different instructions are independent and stateless, \verb+load_instruction_at+ reads from a special system account that stores previous instruction data.
Wormhole verifies the \sym{secp256k} signature from prior payment transactions and then processes the payment.
However, the contract did not \emph{verify} that this account was the intended system account, allowing an attacker to provide counterfeit \sym{secp256k} instruction data to the vulnerable \verb+load_instruction_at+ call.
To mitigate this vulnerability, Wormhole should have checked that the public key of the given account matched the expected system-account key.

This vulnerability is beyond the scope of generic oracles.
VRust~\cite{Cui2022_nm} and \fds~\cite{smolka_2023_fuzz} implement special oracles tailored to this contract's vulnerability only.
The Wormhole function properly validates relevant owner and signer fields except one system-account key; therefore, the generic oracles of VRust, \fds, and \tool do not detect this vulnerability.

At both bytecode and source-code level, a system account is structurally indistinguishable from other accounts.
Without explicit key check, there is no indicator for automated tools that a user account might be unintended.
This is also true at source-code level.
The only hint that a system account is intended can be found in a comment, not in the code itself.
Therefore, generic oracles of automated bytecode or source-code analysis cannot detect this bug.
An analysis that includes comments might be possible now, but also requires at least open-source contracts.

\section{Conclusion}
\tool is the first efficient and practical symbolic execution framework for \ssc.
Based on a large-scale evaluation, we show that \tool significantly outperforms previous approaches.
\tool operates directly on bytecode contracts, supports complex Anchor-based specifications, and is carefully optimized to reduce false positives.
We identified 374 missing key checks (e.g., owner and signer checks) and 100 arbitrary CPI bugs.
These findings show that \tool is a crucial step toward enabling developers to deploy secure contracts on a rapidly growing blockchain.

\paragraph{Acknowledgements}
This work has been partially funded by the Deutsche Forschungsgemeinschaft (DFG, German Research Foundation)---SFB 1119 (CROSSING) 236615297 within project T1, EXC 2092 (CASA) 39078197---, the European Union (ERC, CONSEC, No. 101042266, and Horizon 2020 R\&I, DYNABIC, No. 101070455).
The views and opinions expressed are those of the authors only and do not necessarily reflect those of the European Union or the European Research Council Executive Agency.
Neither the European Union nor the granting authority can be held responsible for them.

{
    \setlength\bibitemsep{4pt plus 2pt minus 4pt}%
        \printbibliography
}
\end{document}